\documentclass[aps,prf,preprint,superscriptaddress]{revtex4-2}

\usepackage{graphicx} 
\usepackage{physics}
\usepackage{xcolor}
\usepackage{multirow}
\usepackage{cleveref}
\usepackage{placeins}
\usepackage{subcaption}
\usepackage{amsmath,amsfonts,amssymb,stackengine,graphicx}
\usepackage{mathtools}

\usepackage{soul}

\newcommand{\C}{\vb{C}}
\newcommand{\I}{\vb{I}}

\renewcommand{\u}{\vb{u}}
\newcommand{\half}{\tfrac{1}{2}}

\newcommand{\change}[1]{{\color{black}{#1}}} 
\newcommand{\delete}[1]{}

\begin{document}

\title{Localised Arrowheads: The building blocks of elastic turbulence in rectilinear, sheared polymer flows}

\author{Theo A. Lewy}
\email{tal43@cam.ac.uk}
\affiliation{Department of Applied Mathematics and Theoretical Physics, University of Cambridge, Cambridge, CB3 0WA, United Kingdom}

\author{Rich R. Kerswell}
\affiliation{Department of Applied Mathematics and Theoretical Physics, University of Cambridge, Cambridge, CB3 0WA, United Kingdom}

%
%
\begin{abstract}

Pressure-driven flow of a dilute polymer solution has been numerically observed to possess a form of  elastic turbulence which is organised around the interactions of localised versions of 2-dimensional `arrowhead' travelling waves (Page et al. {\em Phys. Rev. Lett.} {\bf 125}, 154501, 2020).
As a step to confirming this theoretically, we identify spanwise-localised arrowhead travelling waves by tracking  a symmetry-breaking bifurcation of the known spanwise-invariant (2D) arrowhead to a spanwise-periodic state and then discovering a secondary modulational instability to a spanwise-localised travelling wave. Spanwise-symmetric and asymmetric localised arrowheads exist with the latter having a phase speed slightly inclined to the streamwise direction. Computations capture the flow randomly switching between spanwise local and global arrowhead states in a streamwise-restricted domain, suggesting they form the building blocks of the chaos. Splitting events are also seen, in which a single localised state spawns multiple arrowheads.  However, both cross-shear and spanwise velocities are small suggesting that this elastic turbulence will not be a good mixer.

\end{abstract}

\maketitle

%
%

\paragraph*{Introduction.}\label{intro}

Efficient mixing of fluids allows for enhanced heat transfer and fast chemical reaction rates, both of which have key industrial applications. Elastic instabilities in polymer solutions can promote mixing even at small scales \cite{Burghelea2004} in contrast to inertial instabilities in Newtonian fluids, and consequently elastic turbulence (ET), is touted as a promising way to increase mixing.
Elastic turbulence was first identified in systems with curved streamlines \cite{Groisman_2000}, where transition is caused by linear hoop stress instabilities \cite{Shaqfeh_1996} but was also later found in rectilinear geometries (e.g. experimentally by \cite{Pan_2013} and \cite{Shnapp_2022} in channel flow, and \cite{Bonn_2011} in pipe flow, and numerically by \cite{Berti_2008} and \cite{Berti_2010} in Kolmogorov flow, \cite{Rota_2024, Lellep_2024, Beneitez2024transition} in channel flow, and \cite{Beneitez_2023} in plane-Couette flow) where linear hoop stress instabilities are absent. This suggests that other mechanisms can trigger ET and that possibly multiple types of ET exist.

The centre-mode instability \cite{Garg_Chaudry_Khalid_Shankar_Subramanian_2018, Khalid_2021, Khalid_Chaudhary_Garg_Shankar_Subramanian_2021, Chaudhary_Garg_Subramanian_Shankar_2021} is one such mechanism. This linear instability is 2D, elastic in origin \cite{Buza_2022a}, subcritical \cite{Buza_2022a, Buza_2022b,Page_2020,Dongdong_2021, Morozov_2022, Lewy_2025} and is dynamically connected to exact coherent structures known as `arrowheads' corresponding to the upper branch of solutions \cite{Page_2020}. Significantly, these exist even with vanishing inertia \cite{Berti_2010, Morozov_2022, Buza_2022b, Lellep_2024,Lewy_2025, Nichols2025}. Both recently-observed ET in 3D channel \cite{Lellep_2024} and in 2D Kolmogorov flow \cite{Berti_2008, Berti_2010, Lewy_2025, Nichols2025}  appears organised around localised versions of these arrowhead structures which act like quasi-particles that can collide with each other as well as merge and split \cite{Lewy_2025, Lellep_2024, Nichols2025}.
There is therefore much interest in understanding how these 2D arrowheads sustain themselves  and localise \cite[e.g][]{Morozov2025,Goffin2025, Zhu2025}. Very recently, streamwise-localised versions have been computed \cite{Morozov2025, Zhu2025}: here the focus \delete{here} is on spanwise localisation. 

In this Letter, we isolate both spanwise-periodic and spanwise-localised travelling arrowhead waves in body-forced, rectilinear polymer flow, and demonstrate that spanwise-asymmetric arrowheads can drift in the spanwise direction explaining the collisional dynamics seen in ET. All this is achieved within a spanwise-extended body-forced flow which proves computationally accessible using just a standard workstation  when coupled to periodic boundary conditions in all 3 directions. 
This set-up then facilitates an examination of the likely mixing capability of this form of ET with a preliminary assessment made here.\\


%
%
\paragraph*{Formulation.}\label{models}

We consider 3D viscoelastic Kolmogorov flow in which an Oldroyd-B fluid is driven by a body force in the `streamwise' $\vb{\hat x}$ direction that varies periodically in the `shear' $\vb{\hat y}$ direction, leaving $\vb{\hat z}$ as the `spanwise' direction. The non-dimensional momentum equation is
\begin{equation}
     Re( \frac{\partial \vb{u}}{\partial t} + \vb{u} \cdot \nabla \vb{u})  = - \nabla p + \beta \nabla^2 \vb{u} + \frac{(1-\beta)}{W}\nabla \cdot \vb{C}+ \left(\frac{1+\varepsilon \beta W}{1+\varepsilon W}\right) \cos y \, \vb{\hat x},
     \label{NS}
\end{equation} 
with incompressibility $\nabla \cdot \u = 0$, and the polymer equation is
\begin{equation}
\begin{gathered}\label{gov3:constitutive}
    \frac{\partial{\C}}{\partial t} + \vb u \cdot \nabla \C - \nabla \vb u^T  \cdot \C - \C \cdot \nabla \vb u + \frac{1}{W}(\C - \I) = \varepsilon \nabla ^2 \C
\end{gathered}
\end{equation}
where $\u \coloneqq u\vb{\hat x} + v \vb{\hat y} + w\vb{\hat z}$ is the velocity, 
$\vb{C}$ is the 2nd rank polymer conformation tensor and $p$ is the pressure. 
 All variables are scaled using the laminar peak velocity $U_0$ (which sets the coefficient of the forcing term in (\ref{NS})\,), the total viscosity $\mu = \mu_s + \mu_p$, which is the sum of the solvent and polymer viscosities, respectively, and the length scale $L_0 / 2\pi$ , where $L_0$ is the forcing wavelength. Flow is simulated in a domain of size $[0,L_x]\cross[-\pi, \pi]\cross [-\half L_z, \half L_z]$ with periodic boundary conditions in all directions.
Non-dimensional parameters are the Reynolds number $Re:=\rho U_0 L_0 / 2\pi \mu$, the Weissenberg number $W:=2\pi \lambda U_0 / L_0$, the viscosity ratio $\beta:={\mu_s}/{\mu}$ and the polymer stress diffusion coefficient $\varepsilon := 2\pi \delta/U_0 L_0$, where $\rho$ is the density, $\lambda$ is the relaxation time and $\delta$ is the dimensional polymer stress diffusion coefficient. \change{The 1D base flow that varies only in $y$ is then
$$ \vb U = \cos y \, \vb{\hat x}, \quad P=0, \quad C_{xx} = \frac{W^2}{1+\varepsilon W} \left(1 - \frac{\cos 2y}{1+4\varepsilon W}\right)+1,$$
$$ C_{xy} = \frac{-W}{1+\varepsilon W} \sin y, \quad C_{xz}=C_{yz}=0 \quad \& \quad C_{yy} = C_{zz} = 1.$$}
Throughout this work we set $Re=0.5$, $\beta=0.9$, $\varepsilon=10^{-3}$ but vary  $W$, $L_x$ and $L_z$. At the default choice of $L_x=3\pi$, the unidirectional steady base state is linearly stable at all parameters considered \change{(for a wider discussion of the linear stability of Kolmogorov flow see \cite{Lewy_2025})}.

The system is timestepped using a 3rd order, semi-implicit, backward differentiation scheme \cite{Wang_2008} within the spectral Dedalus software \cite{Dedalus} which uses Fourier expansions in all directions. The resolution used is at least $N_x=64 L_x / 3\pi$, $N_y=64$ and $N_z = 16 L_z/\pi$ \change{(see the supplemental material \cite{supp} for a mesh convergence study)}. 
We work with the reflect symmetries
\begin{align}
\mathcal{R}_y:(u,v,w,p, C_{xx},& C_{yy}, C_{zz}, C_{xy},C_{yz},C_{xz})(x,y,z,t) \nonumber \\
&\rightarrow (u,-v,w,p,C_{xx},C_{yy},C_{zz},-C_{xy},-C_{yz},C_{xz})(x,-y,z,t),\\
\& \quad \mathcal{R}_z:(u,v,w,p, C_{xx},& C_{yy}, C_{zz}, C_{xy},C_{yz},C_{xz})(x,y,z,t) \nonumber \\
&\rightarrow (u,v,-w,p,C_{xx},C_{yy},C_{zz},C_{xy},-C_{yz},-C_{xz})(x,y,-z,t)
\end{align}
which, unless otherwise stated, are both enforced throughout by zeroing either even or odd modes of each field as appropriate. 
Newton methods coupled with GMRES used standardly in polymer-free fluid dynamics \cite[e.g.][]{Chantry2014} unfortunately proved intractable here for reasons not completely clear and so we had to resort to simple time-stepping to identify coherent structures. This meant only attractors could be approached and necessitated a more careful sweep of parameter space to find supercritical bifurcations. These allowed a continuous path of attracting states  to be tracked across bifurcations. Luckily that proved sufficient to at least discover spanwise-localised travelling waves: stable spanwise- {\em and} streamwise-localised waves were not found.\\


%
%

\begin{figure}
\includegraphics[width=0.95\linewidth]{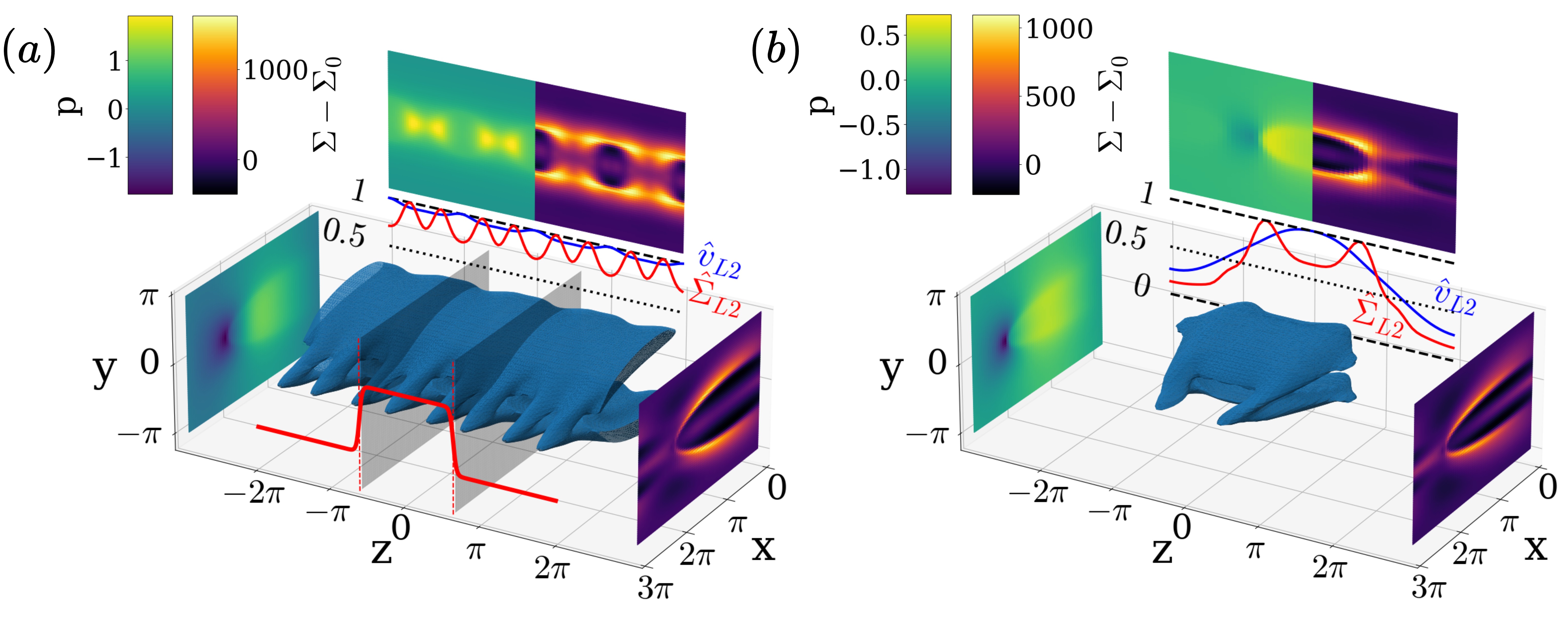}
\caption{Arrowhead solutions at $W=20, L_x=3\pi$ and $L_z=4\pi$. (a) Spanwise-periodic solution with wavelength $\pi$. Blue 3D structure shows the $\Sigma-\Sigma_0 = 450$ isocontour (where $\Sigma:= tr({\bf C})$ and $\Sigma_0$ is the base state value), with surrounding coloured slices showing $p$ ($z<0$) and $\Sigma-\Sigma_0$ ($z>0$) at midplanes where $x=3\pi/2$ (back) and $z=0$ (left and right). The arrowhead structure is shown on the back panel via $\hat \Sigma_{L2}(z)$ (red) and $\hat v_{L2}(z)$ (blue). \change{See \cite{supp} for separate plots of the back panel.} The windowing function $W_f(z;a,b)$ is shown in red at the front when $a=\pi/2$, $b=\pi/4$, with transparent black planes showing where the periodic solution is truncated by the window. (b) Spanwise-localised solution.}
\label{spanwise-solutions}
\end{figure}

{\em Results.} A spanwise-periodic 3D arrowhead was first found by embedding a 2D arrowhead into the 3D system with $L_z=\pi$. While the arrowhead is stable within the 2D system, it is linearly unstable in the 3D system, as is the case in channel flow \cite{Lellep_2023}. White-noise was added to the $u, C_{xx}$ and $C_{xy}$ components of the embedded 2D arrowhead, with amplitude $10^{-2}$ of the maximum base state at $W=20$. Upon simulating to finite amplitude, a stable spanwise-periodic travelling-wave solution was reached: Fig.~\ref{spanwise-solutions}\change{a} shows four spanwise wavelengths. 
The amplitude of the solution across the span is measured in terms of the $L_2$ norm over $x$ and $y$ of the deviation away from the base state, i.e. $\phi_{L2}(z) \coloneqq [\,\int(\phi - \phi_0)^2 dxdy /( 2\pi L_x)\,]^{1/2}$ for a field $\phi$ where $\phi_0$ is its value for the base state, which is then  normalised to give $\hat \phi_{L2}(z) \coloneqq \phi_{L2}(z) / \max(\phi_{L2}(z)) \in [0,1]$. Small values of $\hat \phi_{L2}(z)$ identify regions of the flow that are close to the base state. 

%
%

%
%
\begin{figure}
\includegraphics[width=0.99\linewidth]{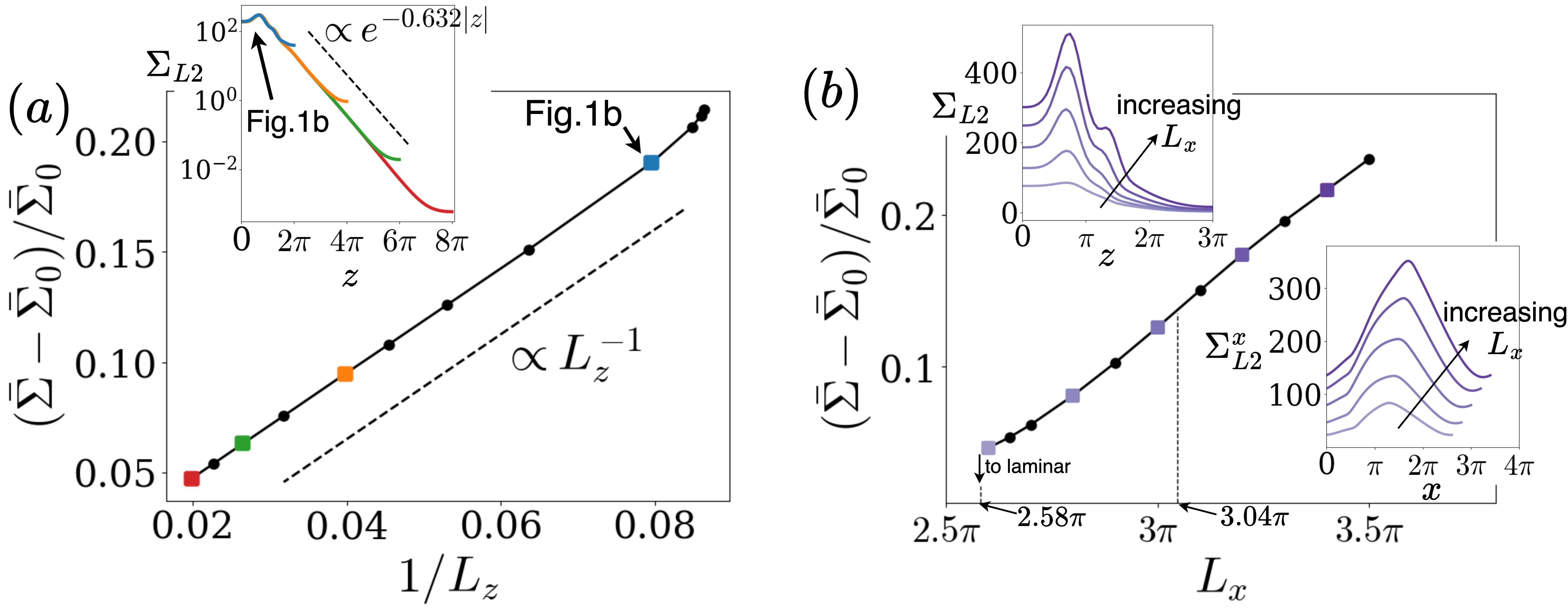}
\caption{(a) The solution of Fig.~\ref{spanwise-solutions}b with $W=20, L_x=3\pi$ continued in $L_z$, with obtained states marked by symbols. Linear dependence at small $1/L_z$ \change{is consistent with} \delete{demonstrates} spanwise localisation, while the branch is lost at larger $1/L_z$. Inset shows $\Sigma_{L2}(z)$ for the\delete{se localised} solutions when $L_z=4\pi,8\pi,12\pi$ and $16\pi$ (marked by coloured squares in (a)) \change{and confirms localisation}. Only $z>0$ is shown as $\Sigma_{L2}(z)$ is even in $z$. (b)~The spanwise-localised solution with $W=20, L_z=6\pi$ continued in $L_x$, with the insets showing  $\Sigma_{L2}(z)$ and $\Sigma^x_{L2}(x) \coloneqq (\int(\Sigma - \Sigma_0)^2 dydz / L_yL_z)^{1/2}$ at points marked by squares. The base state is linearly unstable for $L_x>3.04\pi$. These show that the solution of Fig.~\ref{spanwise-solutions}b is spanwise localised, but not streamwise-localised.}
\label{localising}
\end{figure}

\paragraph*{Spanwise-localisation.} To obtain a spanwise-localised arrowhead, a `windowing' technique was used in which each variable of the spanwise-periodic arrowhead is multiplied by the windowing function $W_f(z;a,b) := f(z)f(-z)$ where $f(z):=\tfrac{1}{2}(1+\tanh[6(a-z)/b + 3]$ \cite{Gibson_2014,Brand_2014} and then time-stepped forward. This windowing function satisfies $0.995<W_f<1$ for $|z|<a$ and $0<W_f<0.005$ for $|z|>a+b$, with $W_f\rightarrow 0$ as $|z|\rightarrow\infty$. This was applied to 4 periods of the spanwise-periodic arrowhead with wavelength $\pi$ in a domain of $L_z=4\pi$, translated in $z$ so that $\hat \Sigma_{L2}$ (where $\Sigma:= tr({\bf C})$) is at a global minima at $z=0$ (see red line on the back panel of Fig.~\ref{spanwise-solutions}a). Setting $a=\pi/2$ and $b=\pi/4$ keeps a whole wavelength of the periodic solution within the window. 
This initial condition evolves under time evolution to a spanwise-localised arrowhead shown in Fig.~\ref{spanwise-solutions}b, with both $\hat \Sigma_{L2}$ and $\hat v_{L2}$ approaching zero far from the plane $z=0$ (see the back panel of Fig.~\ref{spanwise-solutions}b). This state is a travelling wave, as in the spanwise-periodic case. Continuing the solution to higher values of $L_z$ in Fig.~\ref{localising}a confirms spanwise localisation, showing that in fact $\Sigma_{L2}(z) \sim \exp(-0.632|z|)$ for $|z|>2\pi$, while the core region of the solution is independent of $L_z$ for $|z|<2\pi$. \change{Continuation was performed by first stretching (contracting) a converged state into a slightly larger (smaller) domain and then converging via timestepping. The metric $(\bar \Sigma - \bar \Sigma_0) /\bar \Sigma_0$ was used in all bifurcation plots, with a bar denoting a volume average.}

When the spanwise-localised arrowhead in a domain with $L_z=8\pi$ is stretched 50\% to fill a $L_z=12\pi$ domain, or contracted by 50\% to fit into a $L_z=4\pi$ domain, the flow  returns to the original localised state when evolved in time, suggesting that there is a unique stable spanwise-localised state at $L_x=3\pi$. Fig.~\ref{localising}b shows that this state can be continued over $L_x$ with the state no longer stable at $L_x \approx 3.55\pi$ (giving rise to a time-dependent global state) or at $L_x= 2.58\pi$ where there appears to be a subcritical Hopf bifurcation. As $L_x$ increases along this branch, the arrowhead shows no tendency to localise in the streamwise direction (see Fig.~\ref{localising}b right inset) indicating that a streamwise- and spanwise-localised arrowhead is at least one more bifurcation away.


%
%
\begin{figure*}[h!]
\includegraphics[width=0.99\linewidth]{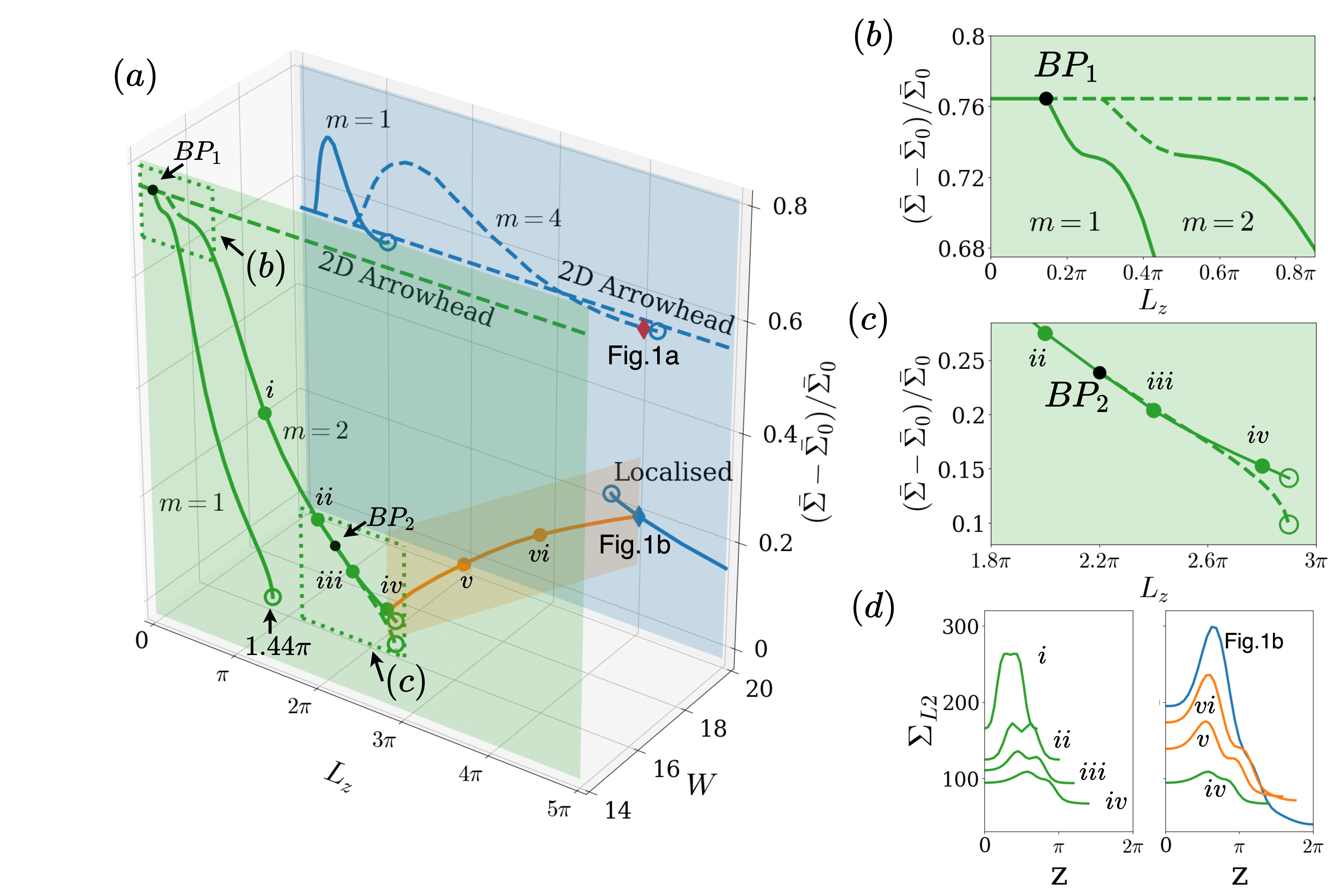}
\caption{(a) The bifurcation plot of $(\bar\Sigma-\bar\Sigma_0) /\bar\Sigma_0$ \delete{(where a bar denotes a volume average)} as $L_z$ and $W$ vary for $Re=0.5, \varepsilon=10^{-3}, \beta=0.9, L_x=3\pi$. Solid (dashed) lines show stable (unstable) solutions. Green curves denotes solutions where $W=14$, blue when $W=20$, and intermediate $W$ are shown in orange. Solution branches are annotated, with `2D arrowhead' denoting spanwise-invariant arrowheads, `$m=(\cdot)$' denoting spanwise-periodic arrowheads with $m$ wavelengths in the domain, and `localised' denoting the branch that localises at large $L_z$ (as in Fig.~\ref{localising}a). The red (blue) diamond at $L_z=4\pi$ correspond to the solutions plotted in Fig.~\ref{spanwise-solutions}a (Fig.~\ref{spanwise-solutions}b). Open circles show where branches could not be continued by time-stepping, implying a bifurcation. \change{The unstable branches were identified from stable solutions within symmetry restricted domains.} We zoom-in on (b) the bifurcation of the spanwise-invariant arrowhead ($BP_1$), and (c) the modulational pitchfork of the $m=2$ branch ($BP_2$). (d) $\Sigma_{L2}(z)$ of solutions $(i)$-$(vi)$ and the localised state of Fig.~\ref{spanwise-solutions}b, showing only $z>0$ as $\Sigma_{L2}$ is even in $z$.}
\label{bifurcation}
\end{figure*}

%
%

\paragraph*{Bifurcations.}

To uncover the bifurcation scenario that generates the spanwise-localised arrowhead (discussed above at $W=20$) by only time-stepping, we had to start from a stable 2D arrowhead in a narrow domain $L_z < 0.14\pi$ at $W=14$. Increasing $L_z$, reveals a symmetry-breaking pitchfork bifurcation at $L_z \approx 0.14\pi$ where a spanwise-periodic travelling wave is born: see BP1 in Fig.~\ref{bifurcation}a where the $\mathcal{R}_y$- and $\mathcal{R}_z$-symmetric solution branch corresponding to one spanwise wavelength (and labelled $m=1$) can be tracked to a saddle-node bifurcation at $L_z \approx 1.44\pi$.  Doubling the width so $L_z=0.28\pi$, however, allows two spanwise wavelengths to be included which are arranged to again be $\mathcal{R}_y$- and $\mathcal{R}_z$-symmetric: this is branch $m=2$ in Fig.~\ref{bifurcation}a. Trivially, because there are two identical waves of wavelength $\half L_z$ across the width and spanwise periodicity, there is also an additional spanwise-shift symmetry: 
\begin{equation}
\mathcal{T}_{z}(\half L_z): \phi(x, y, z) \rightarrow \phi(x, y, z + \half L_z).
\end{equation} 
It is the breaking of this shift symmetry in a modulational pitchfork bifurcation at  $L_z\approx2.2\pi$ ($BP_2$ in Fig.~\ref{bifurcation}c) that gives rise ultimately to a spanwise-localised arrowhead. \change{The two stable branches born at $BP_2$ are translations $\mathcal{T}_{z}(\half L_z)$ of each other, and are indistinguishable in the projection used.} However at $W=14$, this new branch extends up to only $L_z\approx 2.9\pi$ before it is lost, presumably due to a subcritical bifurcation. To avoid the need to navigate through the unstable branch, the stable solution branch at $(W,L_z)=(14,2.8\pi)$ was instead continued across to $(W,L_z)=(20,4\pi)$  where the spanwise-localised arrowhead had been found by windowing (Fig.~\ref{spanwise-solutions}b) using a simple linear interpolation in both parameters. This is shown as 
the orange curve in Fig.~\ref{bifurcation}a. The spanwise-localised arrowhead can be straightforwardly extended by time-stepping to $L_z=16\pi \approx 1/0.02$: see the inset of Fig.~\ref{localising}a. 

The quantity $\Sigma_{L2}(z)$ of the solutions labelled $(i)$-$(vii)$ in Fig.~\ref{bifurcation}a is plotted in Fig.~\ref{bifurcation}d. States $(i)$ and $(ii)$ are periodic over the $z>0$ half width, while $(iii)$-$(vii)$ no longer possess this property due to the modulational bifurcation which breaks the shift symmetry. This modulational bifurcation scenario is exactly how, for example, streamwise-periodic travelling waves  in pipe flow can become streamwise-localised  \cite{Chantry2014}.

%
%

\paragraph*{Spanwise-drifting arrowheads.}
All arrowheads considered so far are $\mathcal{R}_z$-symmetric and so their phase velocity is in the streamwise direction. We now lift this symmetry to find spanwise-asymmetric arrowhead travelling waves which have a non-zero spanwise component of the phase velocity. This scenario is similar to that of the helical travelling wave solutions of Newtonian pipe flow \cite{Pringle2007}, in which structures can slowly rotate while propagating axially downstream.

\begin{figure}
\includegraphics[width=0.5\linewidth]{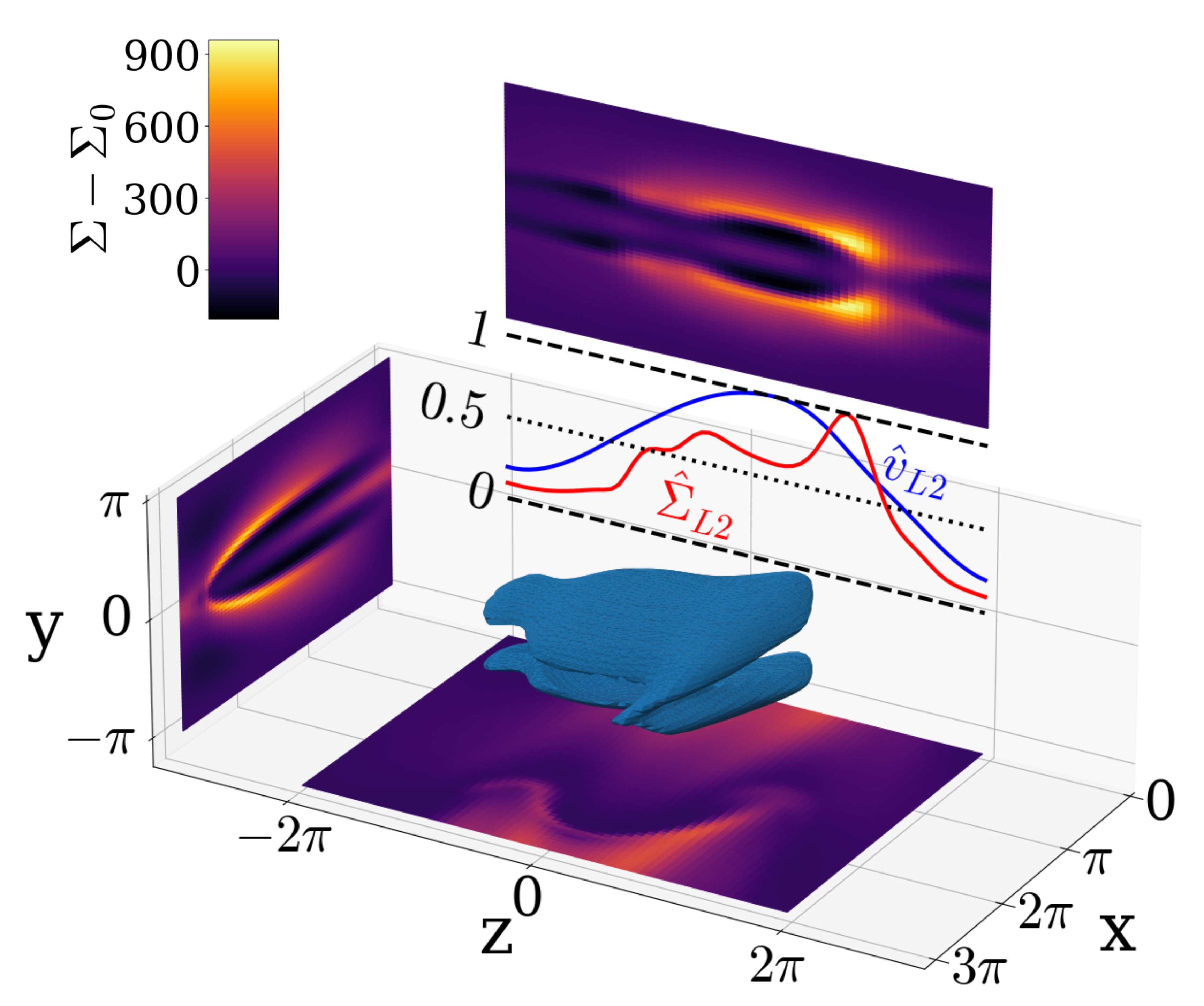}\caption{An asymmetric arrowhead at $W=20, L_z=4\pi$ and $L_x=3\pi$ generated from an intial perturbation with $A=0.1$ (see main text). Isocontours of $\hat \Sigma_{L2}$ and  $\hat v_{L2}$ are as in Fig.~\ref{spanwise-solutions}a. Coloured slices show $\Sigma -\Sigma_0$ midplanes where $x=3\pi/2$ (back), $y=0$ (bottom) and $z=0$ (left). This arrowhead has phase speeds $c_x=0.886$ and $c_z=0.0060$ in the flow and span directions respectively (measured in the frame with no net volume flux in any direction), demonstrating that arrowheads are able to drift in the span direction.}
\label{drifting}
\end{figure}

A simulation was initiated with a spanwise-symmetric, spanwise-localised arrowhead perturbed in the $C_{xx}$ field by the odd function $AC_{xx}\sin(2\pi z/L_z)$, where $A$ is the amplitude of the kick, and converged via time-stepping. The final state was the same for $A=0.05,0.1$ and $0.15$, producing the spanwise-asymmetric arrowhead shown in Fig.~\ref{drifting}. This has phase speeds $c_x=0.886, c_z=0.0060$ in the frame of reference with no net volume flux in any direction. The spanwise drift $c_z$ is orders of magnitude smaller than $c_x$, similar to the slow helical drift of solutions in Newtonian pipe flow \cite{Pringle2007}. Runs with larger perturbations ($A=0.2$) converged to non-localised states. The presence of drifting solutions helps to explain how spanwise-separated arrowheads can collide with each other. \change{These asymmetric arrowheads seem to be born via a subcritical pitchfork from the symmetric arrowhead (see \cite{supp}), suggesting that the symmetric arrowhead loses stability at this bifurcation.}

\change{Kicks in the cross shear direction were also considered. Simulations were initiated with the spanwise-symmetric and spanwise-localised arrowhead and perturbed by $AC_{xx}\sin y$, which is odd in the cross-shear direction with $A=0.1,0.3,0.5$. No new solutions were identified - all simulations eventually returned to the initial symmetric state.}

%

%
%

%

%
%

%
%
\paragraph*{Connection to ET.}

\begin{figure}
\includegraphics[width=0.9\linewidth]{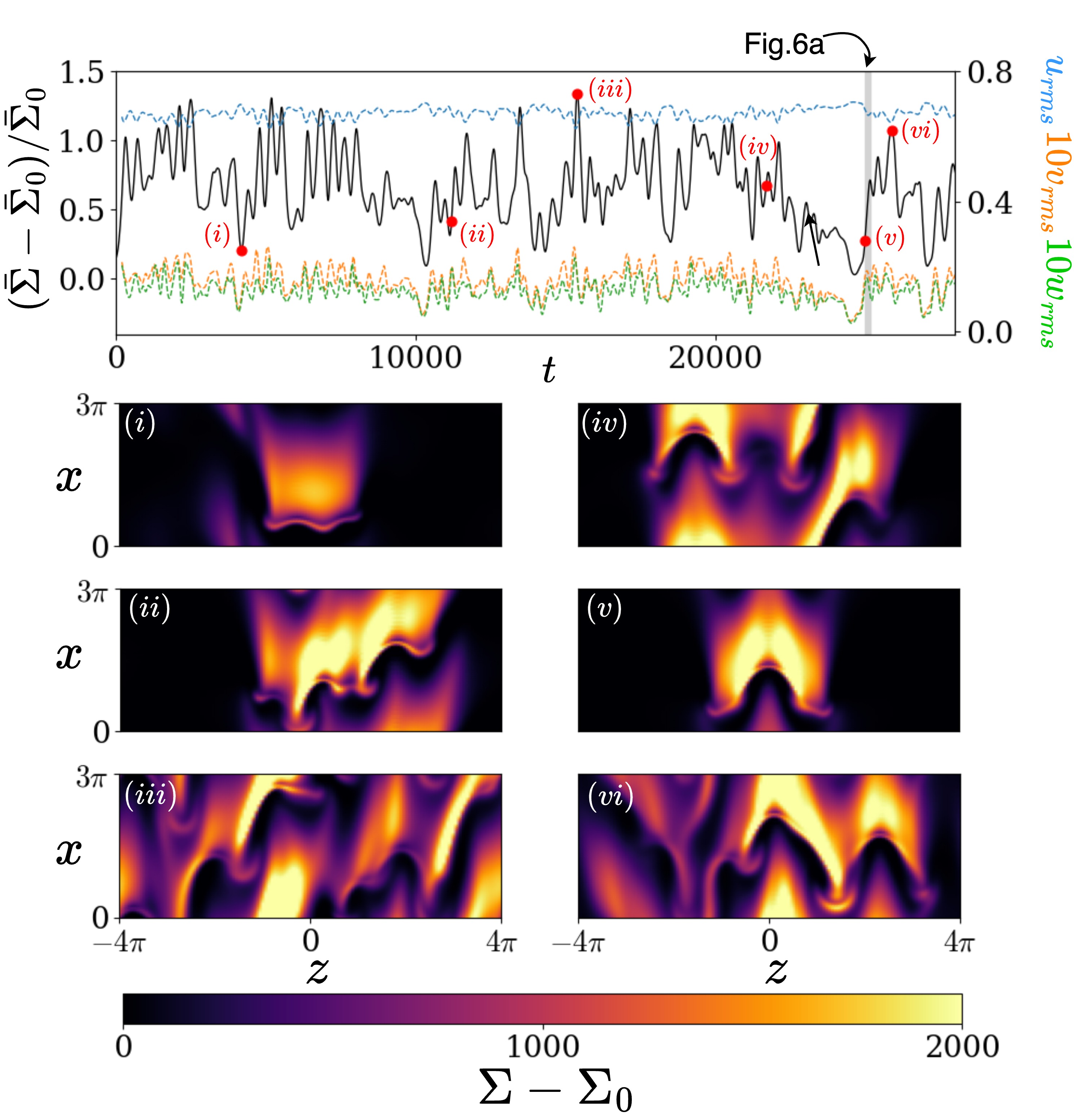}\caption{Elastic turbulence at $W=30$, $\beta=0.95$ in a domain with $L_x=3\pi$, $L_z=8\pi$. The timeseries of $ (\bar\Sigma - \bar \Sigma_0) / \bar \Sigma_0$ and the rms velocity components (top). The narrow grey shaded region corresponds to the time horizon considered in Fig.~\ref{mixing}a. The states $(i)-(vi)$ are marked and $\Sigma-\Sigma_0$ on the $y=0$ midplane is plotted (below). The ET drifts between local states (e.g. $(i)$) and global states (e.g. $(iii)$) over time.}
\label{turbulence}
\end{figure}

The localised structures identified above can be identified  within ET initiated
using the localised arrowhead of Fig.~\ref{spanwise-solutions}b (found at $W=20$, $\beta=0.9$) with the addition of white-noise at $W=30$ and $\beta=0.95$. 
In Fig.~\ref{turbulence} we show how the ET state changes in time. When $(\bar\Sigma-\bar\Sigma_0)/\bar\Sigma_0$ is sufficiently small, the flow is spanwise-localised (see Fig.~\ref{turbulence}$(i), (ii)$ and $(v)$), with states with larger values showing a spanwise-global flow (c.f. $(iii)$ and $(vi)$).  Over time, ET continually transitions between containing local and global structures. 

\begin{figure}
\includegraphics[width=0.9\linewidth]{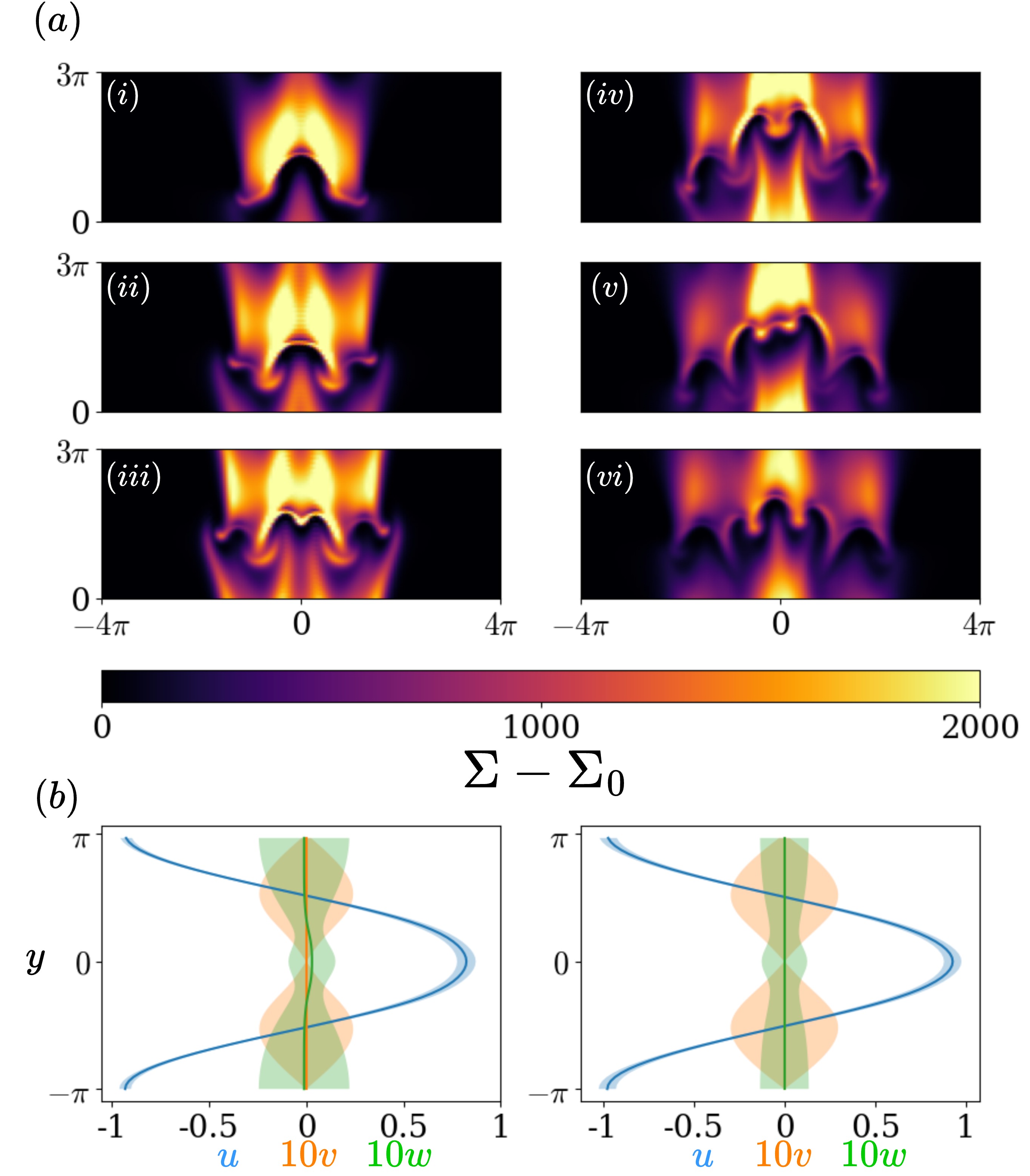}\caption{(a) The single arrowhead of state Fig.~\ref{turbulence}$(v)$ splitting and delocalising. The trace deviation $\Sigma-\Sigma_0$ of the midplane is shown every $T=40$ time units. (b) The stream- and span-averaged velocity components (i.e. $\int \boldsymbol{u} \,dxdz/L_xL_z$) of global state Fig.~\ref{turbulence}$(iii)$ (left) and the spanwise localised state in Fig.~\ref{spanwise-solutions}b (right), where the arrowhead phase speed is in the positive $\boldsymbol{\hat x}$ direction. Solid lines show the mean, while shaded regions show $\pm1$ standard deviation. The cross-shear and spanwise velocities plotted are scaled up by a factor of $10$, but even then have a magnitude much smaller than the streamwise component. }
\label{mixing}
\end{figure}

Both the symmetric and asymmetric arrowheads are visible in the $y=0$ midplane slices shown. In Fig.~\ref{turbulence}$(v)$ a symmetric arrowhead is visible and also discernable in $(iv)$ for $z<0$. The asymmetric arrowhead is prominent in $(ii)$ (with two in the domain) and $(vi)$ (at $z=0$). These look similar to the bottom panel of the asymmetric arrowhead in Fig.~\ref{drifting}. Due to the presence of these localised structures in this subcritical turbulence, it seems that ET is organised around them consistent with prior large-scale simulations of wall-bounded body-forced flow \cite{Lellep_2024, Morozov2025}.
A transition from one symmetric arrowhead into multiple is also observed. In Fig.~\ref{mixing}$a$ we show how state Fig.~\ref{turbulence}$(v)$ evolves over $200\, L_0/2\pi U_0$. The initially symmetric arrowhead $(i)$ grows weak arrowheads on its sides $(ii)$ before pinching in the middle $(iii)$. At this point the structure loses its $\mathcal R_z$ symmetry $(iv)$. The pinched region then grows to form a separate arrowhead $(v)$, $(vi)$. Up to 5 separate arrowheads can be seen in state $(vi)$, and the spanwise extent of the state has increased from the original symmetric state $(i)$. This splitting event provides a pathway to delocalise a given state in ET.

\paragraph*{Mixing.}

The size of the fluctuations in ET is a good first indicator of the mixing potential of the flow. Fig.~\ref{turbulence}(top) shows time series of the root mean square velocity components, $u_{rms}$, $v_{rms}$ and $w_{rms}$ for ET  (with $\phi_{rms}(t) := \sqrt{\int \phi^2dV/2\pi L_xL_z}$). This indicates that the streamwise velocity is over an order of magnitude larger than the other velocity components (time-averaged means are $\langle u_{rms}\rangle_T=0.6708$, $\langle v_{rms}\rangle_T=0.0156$ and $\langle w_{rms}\rangle_T=0.0127$ in the frame with vanishing volume flux in all directions). Turning now to a specific snapshot
- the global state Fig.~\ref{turbulence}$(iii)$ - the mean and standard deviation of the velocity over each plane of constant $y$ are shown on the left-hand side of Fig.~\ref{mixing}b. The cross-shear and spanwise velocities have very small mean, and their standard deviations are generally much smaller than the mean streamwise velocity throughout the flow. The value of $v_{rms}$ and $w_{rms}$ are less than $2\%$ of the mean $u$ on the centreline consistent with wall-bounded simulations \cite{Lellep_2024}. This indicates that very little motion is induced in directions perpendicular to the forcing. The right-hand side of Fig.~\ref{mixing}b shows the same is true for the spanwise localised arrowhead state of Fig.~\ref{spanwise-solutions}b, where the mean cross-shear and span velocities vanish exactly by symmetry. These observations suggest that ET based around these arrowhead states will be a poor mixer.\\

%
%

%
%


%
%

\paragraph*{Discussion.}

We have identified spanwise-periodic arrowheads, as well as symmetric and asymmetric spanwise-localised arrowheads which underpin the dynamics of ET in a body-forced, 3-dimensional, rectilinear, polymer flow with periodic boundary conditions (viscoelastic 3D Kolmogorov flow). The asymmetric spanwise-localised arrowheads drift in the spanwise direction and can therefore initiate the collisional dynamics seen in spanwise-extended domains. 
These spanwise-localised arrowhead waves are clearly visible for (subcritical) Weissenberg numbers where the base state is linearly stable and are seen to split into multiple, possibly drifting arrowheads, causing delocalisation of ET. This global state then at some point collapses back to the localised form to repeat the cycle. Both the global states and the localised states have small velocities perpendicular to the forcing, suggesting that ET will be a poor mixer. That even the global state is poor gives little hope that this conclusion would change for higher (supercritical) Weissenberg numbers where the base state is unstable ensuring that ET is persistently global. 

The stable symmetric spanwise-localised arrowhead found here appears unique for a given streamwise wavelength and has one spanwise wavelength of the associated spanwise periodic arrowhead at its core. Surely many more unstable versions exist with more wavelengths included in their heart borne from similar modulational pitchfork bifurcations. Similarly both streamwise- and spanwise-localised arrowheads must exist but presumably are unstable and so beyond our admittedly weak time-stepping grasp here. And finally, a fully-localised arrowhead is entirely conceivable in this triply-periodic set-up. Just like in our  spanwise-focussed study here, this could arise through a modulational instability in the cross-shear direction but with the added complication that the underlying state is cross-shear-periodic  and so a more involved modulational Floquet analysis is required to capture this.

\change{Finally, it is worth remarking that localised coherent structures have also been suggested as building blocks of hoop-stress-driven elastic turbulence. Axially-localised `diwhirl' structures were observed experimentally \cite{groisman1997} in Taylor-Couette flow and then isolated computationally \cite{Kumar2000,KUMAR_GRAHAM_2001}. Subsequently it has been suggested that interactions of these localised diwhirls with elastic waves sustain this type of ET \cite{Song_Liu_Lu_Khomami_2022,Song2023}.} 

In terms of future work, the triply-periodic system studied here with only one forcing wavelength included is a particularly accessible flow to study arrowhead-centric ET as wall processes are of secondary importance.  For example, it would be well within reach using a single workstation to add a passive scalar  or even an active field to study the mixing characteristics of this flow. From a methodology perspective, this 3D viscoelastic flow is also an excellent test bed to improve or even fix the sophisticated dynamical systems tools (e.g. GMRES) for capturing unstable exact solutions which have proved so effective in Newtonian flows. We hope to report on at least one of these in the near future.


\FloatBarrier

%
%
\bibliography{9.bibliography} 
\FloatBarrier 




\end{document}